# UCLog+ : A Security Data Management System for Correlating Alerts, Incidents, and Raw Data From Remote Logs

William Yurcik[4]   Cristina Abad[1,2,4]   Ragib Hasan[2,4]   Moazzam Saleem[2]   Shyama Sridharan[3]

[1]Escuela Superior Politécnica del Litoral (ESPOL)

[2]Department of Computer Science
[3]Department of Electrical and Computer Engineering
[4]National Center for Supercomputing Applications (NCSA)
University of Illinois at Urbana-Champaign

*byurcik@ncsa.uiuc.edu, cabad@fiec.espol.edu.ec, {rhasan,msaleem2,sridharan}@uiuc.edu*

## Abstract

*Source data for computer network security analysis takes different forms (alerts, incidents, logs) and each source may be voluminous. Due to the challenge this presents for data management, this has often lead to security "stovepipe" operations which focus primarily on a small number of data sources for analysis with little or no automated correlation between data sources (although correlation may be done manually). We seek to address this systemic problem.*

*In previous work we developed a unified correlated logging system (UCLog) that automatically processes alerts from different devices. We take this work one step further by presenting the architecture and applications of UCLog+ which adds the new capability to correlate between alerts and incidents and raw data located on remote logs. UCLog+ can be used for forensic analysis including queries and report generation but more importantly it can be used for near-real-time situational awareness of attack patterns in progress. The system, implemented with open source tools, can also be a repository for secure information sharing by different organizations.*

## 1. Introduction

For security monitoring and forensics, organizations currently use a mix of commercial, open source, and homegrown system scripts to generate alerts. In addition, many organizations have incident response teams that investigate and document security events often using an automated incident database system for indexed retrieval. Lastly, there are many logs, which store raw data about device and network status. Typically these three sources of information relevant to security (alerts, incidents, raw data logs) are handled in different systems and any correlation between these sources is accomplished only with manual intervention (both complex and tedious).

In previous work we developed a unified correlated logging system (UCLog) that automatically processes alerts from different log sources [7]. We performed a proof of concept that showed automated log correlation of alerts is both feasible and beneficial [1].

In this paper we present UCLog+, which is an incrementally improved system which adds the capability to correlate beyond alert logs to also include incidents documented by an incident response team and raw data in common network logs (syslogs, NetFlows). The goal is automated integrated access between alerts, incidents, and raw data.

The remainder of this paper is organized as follows: Section 2 surveys related work. Section 3 discusses available data sources. Section 4 presents the architecture of UCLog+. We present experimental results in Section 5. We end with a summary and conclusions in Section 6.

## 2. Related Work

Several tools or methods to increase the efficiency of security operations in the management of security-related data have been proposed. MieLog [12] provides visualization and statistical analysis tools for specific logs. NVisionIP is a situational awareness visualization tool for NetFlows raw data logs that can be extended to integrate other data sources [5]. Fisk et al. developed a set of tools that provide SQL-like manipulation of streaming network data from different sources [3]. Lee and Stolfo proposed using data mining to correlate IDS alerts and improve alert accuracy [6]. Sah [10] describes a database for large-scale archival.

Computer security incident databases have been proposed to integrate different methods of access and query. In particular, Ohio-State University developed two incident databases: IDB and SITAR. IDB automatically inserts simple alerts (with minimal information), and SITAR is used for manually storing more extensive incident information [2]. UCLog+ differs from these in its design choices and functionality. The databases from Ohio-State do not consider raw data logs (other than email alerts) and these systems are not well documented, making it hard to learn from their experience.

UCLog+ complements these existing solutions by providing a central repository for data of different types -- alerts, incidents, and raw data can be correlated from the same interface.

## 3. Security Data Sources

There are many potential data sources for security analysis. Since each computer network environment is unique (in terms of assets, threats, and vulnerabilities) we must consider the widest range of potential data sources in order to provide maximum flexibility for organizations to tailor specific solutions.

The most basic issue is the form of data sources to consider. At the lowest level kernel events provide state change information and deal with kernel entities such as file descriptors and process IDs. However, data management of kernel events does not scale except for all but the smallest networks. Also humans who perform security analysis are interested in higher-level state such as files, process names, and connections over specified time intervals. Examples of requirements elicited from security engineers include the top processes, top connections, and file modifications in the last hour; attacks which match known signatures; and anomalous activity exhibited by rare or new behavior.

The need for high-level situational awareness makes alerts from detectors appropriate for one form of data source. Alerts can be generated by firewalls, intrusion detection systems (IDSs) such as Snort [9] or Bro [8], file system integrity checkers like Tripwire [4], or alert generation tools for applications like RedAlert [11]. Large networked systems usually run a mixture of different alert generators.

A second form of security source data we consider is incident data documented from response teams investigating events. These events may be initiated from alarms or from human requests. Sometimes this information is referred to as meta-alarms since humans document their analysis of multiple alarms (multiple machines compromised by the same attack, multiple attacks on the network, etc.). Incident data may be stored in writing, Email messages, flat computer files, a database, or sometimes not formally stored at all. For instance, until recently the events reported to and investigated by the CERT/CC were stored only in Email format. Incident data is a rich source of security information since it often includes insight from human analysis over time. However, historical incident data has been difficult to integrate into security analysis since it is hard to parse from free-form writing in different formats.

The third and last form of security source data we consider are raw data logs. Examples include NetFlows and application-specific syslogs (e.g. web, Kerberos, dhcp, snmp, etc.). NetFlows are a network based log that can be generated by routers (e.g. Cisco) or specialized open source software (e.g. Argus) documenting traffic flows passing through an observation point. These raw data log files are typically flat files (not organized in any way) that record status events from sources as they occur in near-real-time. The status events can be in different granularities such as low (record only important events) to high (record everything).

For this work we converge on these three forms of source data: (1) alarms, (2) incidents, and (3) raw data logs). There are obviously more than three forms of source data for security analysis but we posit these three forms are commonly implemented by organizations. Our goal is to create a system to access these three different forms of source data in a central repository that allows for automated correlation and flexible query ability. While it can be argued that a security data management system is not needed since data in these three forms is already available for analysis, we find from personal experience that manually correlating information between these different forms is painful to accomplish such only attempted in the extreme cases.[1]

---

[1] The first author has experience as Head of Security Operations and Incident Response at NCSA 2002-2003.

## 4. UCLog+ Architecture/Implementation

The concept behind the UCLog+ system architecture is to store alerts and incident records in a central database while launching search commands to large raw data files using encrypted ssh connections. This allows raw data logs to remain where they are located without moving their entire contents over the network. The records in the database can be queried directly and scripts are created for search commands against raw data files whose results will be delivered back to the database and back to the user GUI. The end result is transparent awareness of events for security analysis without the user having to manually manipulate alerts, incidents, or raw data logs.

The UCLog+ system consists of several independent and cooperating elements. Figure 1 shows the relationship between these elements:

- Cron jobs to process the incoming email alerts
- Perl scripts parse the alerts and automatically insert the necessary records in the database.
- A PostgreSQL database for incident storage
- Pgpsql procedures for specific database operations
- PHP scripts for the web-based GUI
- Flat NetFlow logs stored on a different server
- Perl scripts for correlation with flat NetFlows

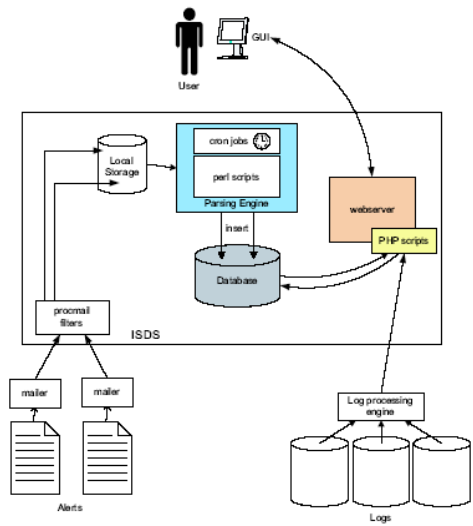

**Figure 1.** UCLog+ Architecture

The database consists of several tables: incidents, types, Emails and Hosts. The *types* table contains incident or alert types (e.g. scan, password attack) information; it can be as specific as the alerts allow.

The *hosts* table stores information about internal/external hosts involved in one or more incidents. The *Emails* table stores emails detailing the incidents, including original alerts sent in emails to the system and non-standard follow-up emails between Security Engineers. Finally, the *incidents* table is the main table, containing one record per incident involving a host; comments can be added to help revisit the incidents in the future. The table creation script is shown in Table 1.

```
CREATE SEQUENCE seq_hostid;
CREATE SEQUENCE seq_incidentid;
CREATE SEQUENCE seq_typeid;
CREATE SEQUENCE seq_emailid;
CREATE TABLE hosts (
    hostid INTEGER PRIMARY KEY
        DEFAULT nextval('seq_hostid'),
    name VARCHAR(30) UNIQUE NOT NULL
        CONSTRAINT invalid_host_name
        CHECK(name LIKE '%.%.%.%'),
    ip INET,
    owner_name VARCHAR(35),
    owner_email VARCHAR(35)
);
CREATE TABLE emails (
    emailid INTEGER PRIMARY KEY
        DEFAULT nextval('seq_emailid'),
    date DATE NOT NULL,
    source TEXT,
    comments TEXT
);
CREATE TABLE types (
    typeid INTEGER PRIMARY KEY
        DEFAULT nextval('seq_typeid'),
    name VARCHAR(25) UNIQUE NOT NULL,
    description VARCHAR(256)
);
CREATE TABLE incidents (
    incidentid INTEGER PRIMARY KEY
        DEFAULT nextval('seq_incidentid'),
    date TIMESTAMP NOT NULL,
    host INTEGER REFERENCES hosts (hostid),
    type INTEGER REFERENCES types (typeid),
    email INTEGER REFERENCES emails (emailid),
    comments TEXT
);
```

**Table 1.** Table Creation Script

Storing and organizing alerts in a relational database adds value by allowing complex queries that would otherwise be time challenged due to parsing of flat files containing alerts or incidents. Common pre-defined queries and customizable queries can be accessed from the web interface. A free-form SQL query interface is provided for increased flexibility. The query results are shown in HTML tables and can be downloaded as plain text files (delimited by TABs) if needed. For example, the SQL statement for a pre-defined query is shown in Table 2.

```
SELECT incidents.date, hosts.name, types.description
FROM incidents,hosts,types,emails
WHERE host=hostid AND type=typeid AND
email=emailid;
```
**Table 2.** Pre-Defined SQL Query

Several PHP scripts generate HTML pages that provide a user interface to the database data. The PostgreSQL database can also be queried via command-line, but the HTML interface is more user-friendly. When viewing the incidents, the user directs queries transparently to the appropriate raw data logs through the web interface. Perl scripts are used to parse the raw logs for a particular host and/or time period and generate smaller and easier to handle files that can be viewed on-line. Other Perl scripts parse the incoming email alerts and insert them into the appropriate tables, performing host name/IP lookup if necessary.

Reports and statistics are also accessible from the web interface. Gnuplot is accessed by PHP scripts to generate graphs showing different useful information obtained by querying the database. For example, a pre-defined report in the system queries the database to obtain a percentage of incidents per day of the week. The SQL query is shown in Table 3.

```
SELECT my_dow(date) AS day,
count(*)/(select count(*)/100.0 from incidents) AS cnt,
extract(DOW FROM date) AS dow
FROM incidents GROUP BY day, dow
ORDER BY dow;
```
**Table 3.** SQL Query for a Report

New types of alerts or alert sources can be added by writing new parsing scripts to process the new types of alerts and insert them into the database. Alerts and incidents from different sources can be correlated with data from raw data logs by matching fields (usually host and/or time period).

Searches against raw data logs launched transparently by the user are cached for later use. This speeds viewing and helps preserve the raw data in case it is aged and deleted from the original repository.

In our current implementation we are able to correlate IDS alerts and other in-house developed alert generators with raw data from NetFlows logs. We are working to include syslog from web servers, main servers, and authentication servers.

### 4.1 GUI Design

Figure 2 shows the main web-based GUI for UCLog+. The system has different views and access control features for administrators and normal users. Administrators can view more sensitive information while the normal users can only access non-critical information regulated by the administrators.

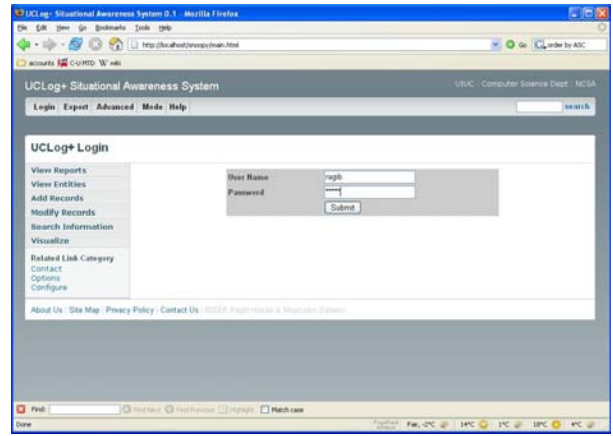

**Figure 2.** UCLog+ Web Interface

Figure 3 shows the main report generation interface for UCLog+. The administrator can select different options to execute different types of queries. Queries can be displayed in tabular or graphical format.

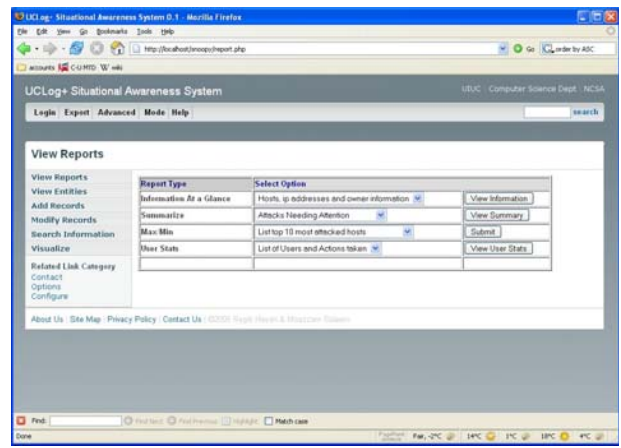

**Figure 3.** UCLog+ Query Interface

Figure 4 shows a sample plot generated by UCLog+ -- the number of attacks per host and the frequency of the type of attacks.

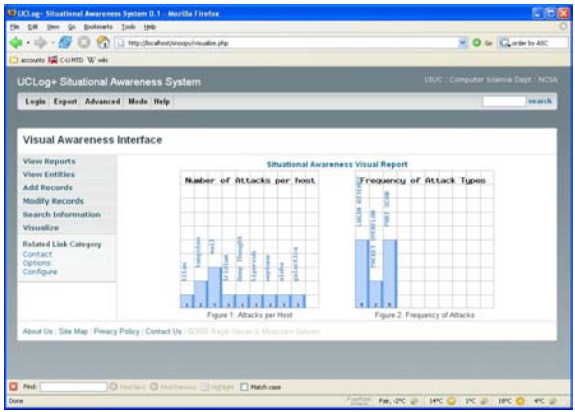

**Figure 4.** UCLog+ Sample Visual Output

## 4.2 System Security

While the motivation for developing this system is to enhance security, it does present several security issues -- in particular: audit, unauthorized access and alert email validity. All modifications to database data should be logged. Access control policies and authentication are needed. Email messages sent to the server containing alerts to be included in the database must be signed to ensure that they come from an authorized source. For now we have a simple policy:

*Allow all or deny all, depending on the user. A login and password is needed for viewing/ modifying data. SSL is used to secure HTTP communication. SSH is used for secure connections between the database and logs.*

## 5. Preliminary Results

Preliminary results report that UCLog+ intuitively enables correlations between alerts, incidents, and raw data. The flexibility of the SQL queries adds value by obtaining information that would otherwise not be obtained due to effort.

While in the past we have demonstrated correlation between alerts, UCLog+ enables new correlations between incidents inside the database and external raw data logs. These correlations are typically focused on date/time and host IP address. When browsing through an incident report in the system, the user can click on a "show NetFlows" link which calls a PHP script that parses the appropriate NetFlows raw data log located in a different server available via an SSH connection over the Internet. For example, if an alert is generated for a particular IP address that is scanning – that IP address is noted (blacklisted) and at the click of mouse a query can be launched to NetFlows raw data logs requesting all IP addresses the blacklisted IP address has scanned during a given period (hour/day/week). The resulting query is transparently launched and executed remotely on the raw data log file and the IP address output is shown to the user in an HTML report. If one of the scan targets has appeared in one or more posterior alerts, a security engineer may again access the corresponding NetFlow raw data logs through the UCLog+ interface to learn additional information such as the direction of flows (source/destination), port usage, etc. Lastly, if a security engineers identifies an attack patterns with specific packet sizes or ports, they can create new alert signatures to feedback to sensors (Closing-the-Loop).

We have implemented UCLog+ using alerts from the BRO intrusion detection system, two years worth of incident records from NCSA's response team, and Netflows raw data logs. After preloading historical data and setting up pipes for continuous loading of new data, UCLog+ was used to correlate information to learn about more about the types of attacks being experienced. We were able to find these interesting trends in attacks targeting NCSA.

**Attack Types:** Of the various attacks on the NCSA server, we can see from Figure 5 that the Denial-of-Service, Brute Force Password Attempts, and Port Scans comprise about 81.3% of all attacks.

**Timing of Attacks:** From Figures 6 and 7, it can be seen that 83% of the attacks occur on weekdays and most of the attacks are in the time range 8am-6pm. This pattern is also mirrored in other attack distributions.

**Attack Patterns:** The attacks from host X follow a fixed pattern. In Figure 14, it can be seen that the attack pattern matches a normal curve with mean 57 and standard deviation 23.76. While in Figure 15, it is observed that attacks form a pattern w.r.t. day and time, statistically it can be seen that the number of attacks on each of the days is about the same -- only standard deviation per day varies.

**Source of Attacks**:
Scans: One-fourth of the scans can be traced to different IP addresses in Netherlands
Password Attacks: One-half of the password attacks from one host. 22% of the remaining from Australia and 11% from a specific internal machine.
Password Attacks: One-half of the password attacks from one host. 22% of the remaining from Australia and 11% from a specific internal machine.
**Trend in Scanning:** Most of the scans have been in the range 141.142.0.x and 141.142.110.x . 41.3% of

scans in the IP address range 141.142.65.x and 141.142.67.x, 13.33% in the range 141.142.2.x, 14.67% in the range 141.142.96.x and 141.142.105.x and the remaining 2.67% in the range beyond 141.142.200.x

**Trends in Password Attacks:** The host *X* seems to be compromised with one of the attacks originating from within our internal network. The attacks pattern appears to be programmed to perform 25 username/password combinations between the time range 2:05am to 2:21am and a 100 username/password combinations between the time range 8:45am to 9:15am (Figure 13).

**Forensics:** A new un-alerted incident was revealed. The alert is of a kind that requires investigation to determine if any damage has occurred (e.g. compromised hosts, etc.). With the UCLog+ system in place, a next logical step is to retrieve previous incidents related to one or more of the involved hosts. The query is shown in Table 4. This query leads to valuable information about the hosts involved (e.g. had been scanned in previous weeks by the host now trying dictionary attacks to brute-force access to it).

| SELECT * FROM incidents, hosts, emails, types WHERE host=hostid AND type=typeid AND email=emailid AND hosts.name='w.x.y.z' ORDER BY incidents.date; |
|---|
| **Table 4.** Tracing Events Between Hosts |

**Preventive Security:** The list of top ten internal hosts (regarding appearance in incidents) can be obtained for careful analysis and patching of each to avoid future incidents. Table 5 shows the SQL query that determines this.

| SELECT hosts.name, hosts.ip, count(incidentid) AS cnt FROM hosts, incidents WHERE hosts.hostid=incidents.host GROUP BY hosts.name, hosts.ip ORDER BY cnt DESC LIMIT 10; |
|---|
| **Table 5.** Query for Top-10 Compromises |

**Finding the Usual Suspects:** If an *external* host is frequently appearing in incidents, careful analysis of these historical incidents will help understand why. If after analysis we determine that an external host is responsible for these incidents, we may notify the ISP of the external host or block access to that IP address.

If an *internal* host is frequently appearing in incidents, careful analysis will help understand why this is happening and take preventive measures (e.g. if host is infected with a virus, the host must be ``cleaned''). In our experimental data, it can be seen from Figure 16 that 7 hosts: A,B,C,D,E,F,G stand out as most frequently attacked hosts and therefore administrator's efforts should be directed at taking preventive measures for these hosts.

**Policy Violation:** A list of top ten internal hosts regarding type of anomaly (e.g. DoS) can be generated to investigate possible policy violations performed by the users of those hosts (e.g. illegal file/media sharing) and issue appropriate warnings to those users. Sample SQL code generating this report is shown in Table 6.

| SELECT hosts.name, hosts.ip, count(incidentid) AS cnt FROM hosts, incidents, types WHERE hosts.hostid=incidents.host AND types.name = 'INCBAND' GROUP BY hosts.name, hosts.ip ORDER BY cnt DESC LIMIT 10; |
|---|
| **Table 6.** Query for Policy Violations |

**Report Generation:** Statistical incident information can be generated to include in reports, including percentages and graphs as Shown in Figures 6-13. For example, to evaluate if a new plan to secure the internal network is working, a trend analysis report over time listing the number of incidents per time period for multiple time periods can be displayed. The corresponding query is shown in Table 7.

| SELECT MOD((extract(MONTH FROM date)+12 - extract(MONTH FROM CURRENT_DATE)-1)::integer,12) AS pos, extract(MONTH FROM date) AS mon, count(incidentid) FROM incidents WHERE CURRENT_DATE - date < '365 days' GROUP BY mon ORDER BY pos; |
|---|
| **Table 7.** Trend Analysis Report Generation |

**Attacking the Problem at its Roots:** Identifying recent ``first offenders'' (i.e. internal hosts that appear in an incident for the first time) can help in preventing future problems (e.g. by patching host, installing firewalls or warning users in case of policy violations). Table 8 shows the SQL query for finding recent first offenders.

| SELECT hosts.name, ip, MAX(date) FROM hosts, incidents WHERE hostid=host AND CURRENT_DATE - date < '1 month' GROUP BY date,hosts.name, hosts.ip ORDER BY date DESC LIMIT 10; |
|---|
| **Table 8.** Query for Recent Successful Attacks |

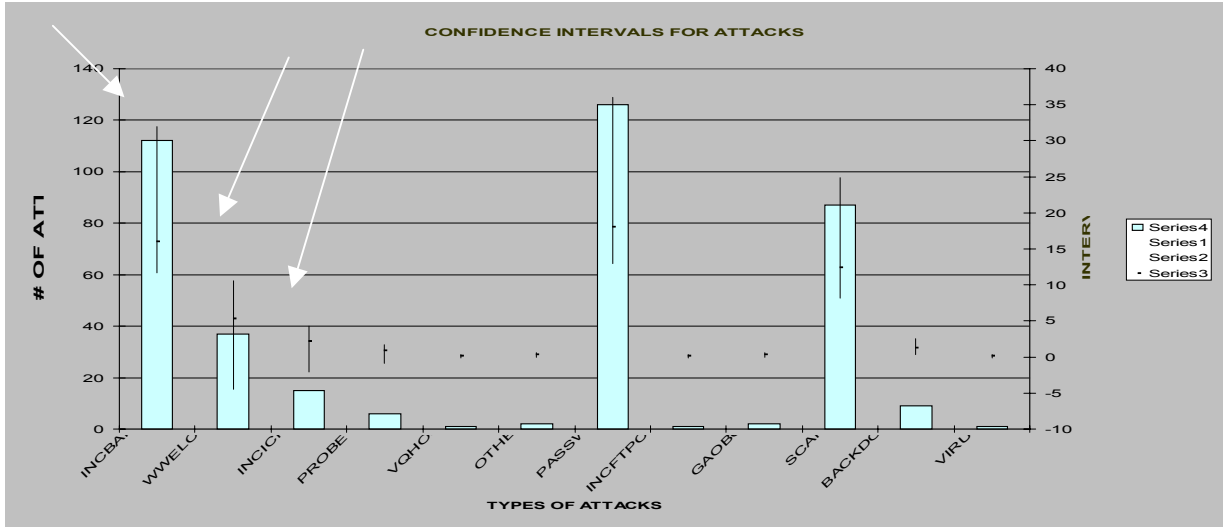

**Figure 5.** Attack Type Distribution

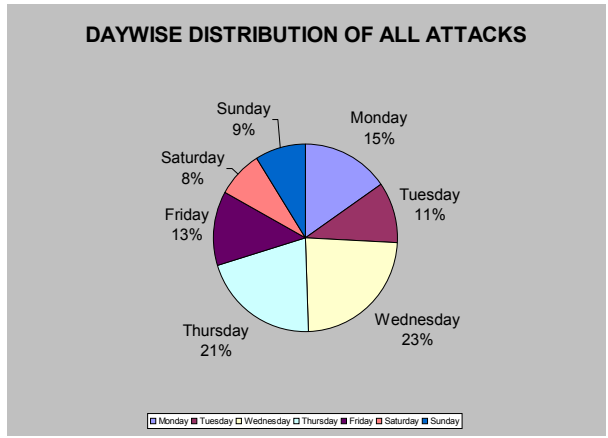

**Figure 6.** Attack Distribution by Day of Week

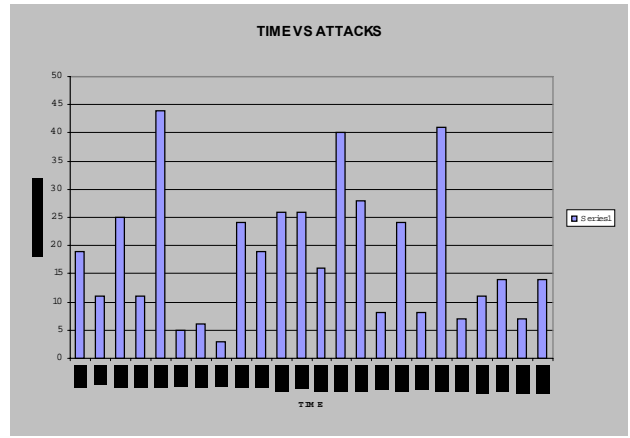

**Figure 7.** Attack Distribution by Time

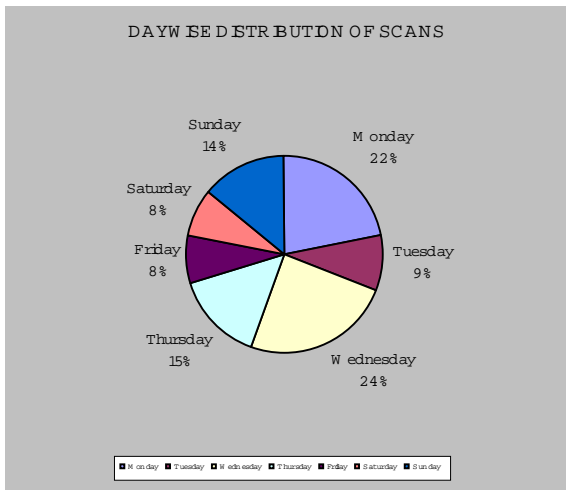

**Figure 8.** Scan Distribution by Day of Week

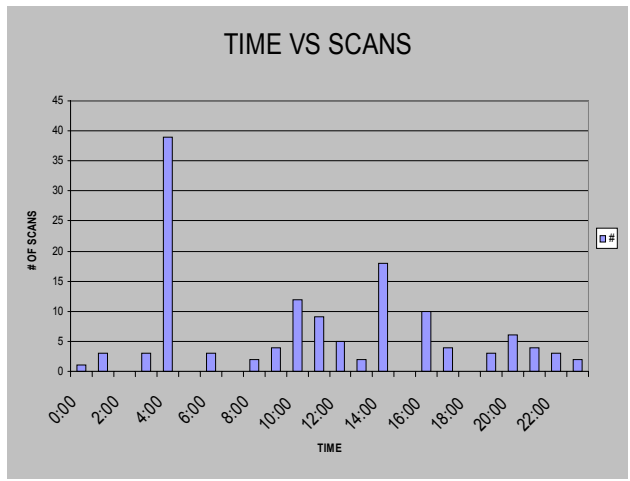

**Figure 9.** Scan Distribution by Time

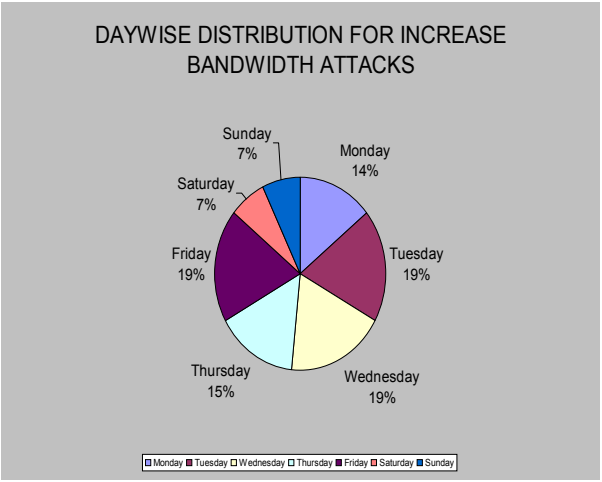

**Figure 10.** DoS Attack Distribution by Day

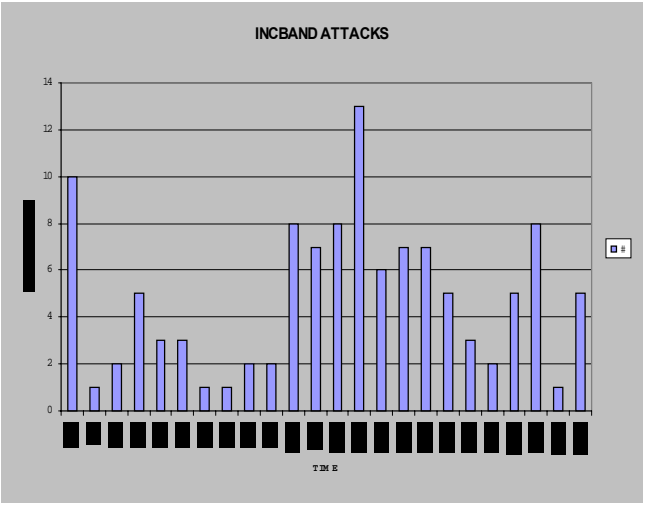

**Figure 11.** DoS Attack Distribution by Time

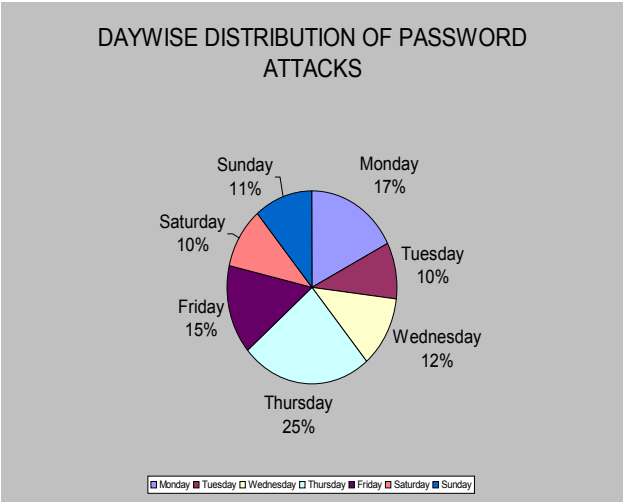

**Figure 12.** Password Attack Distribution by Day

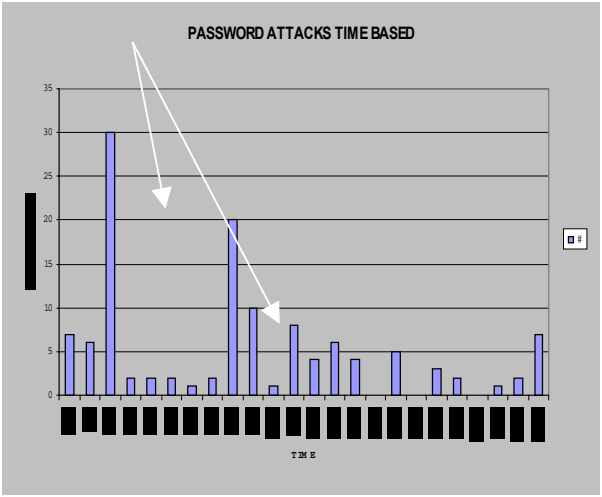

**Figure 13.** Password Attack Distribution by Time

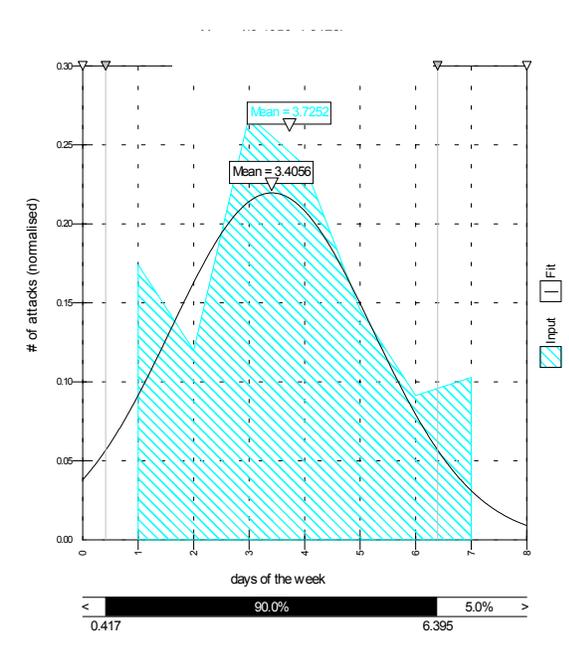
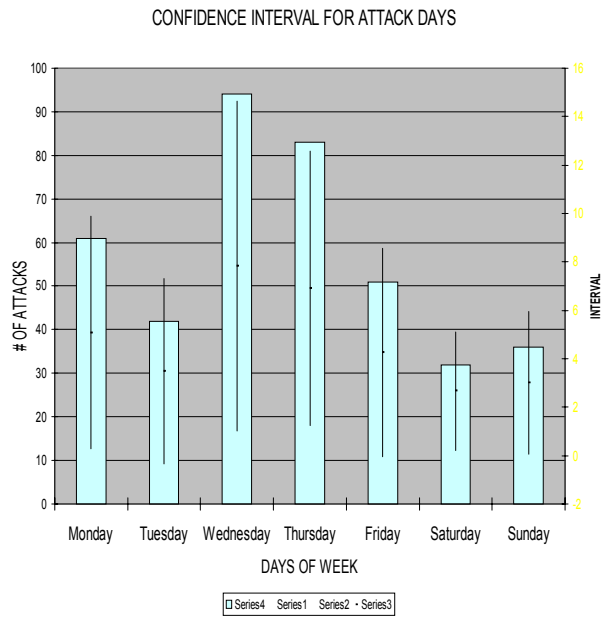

**Figure 14.** Normalized Distribution of Attacks from Host X

**Figure 15.** Confidence Level for Attacks from Host X

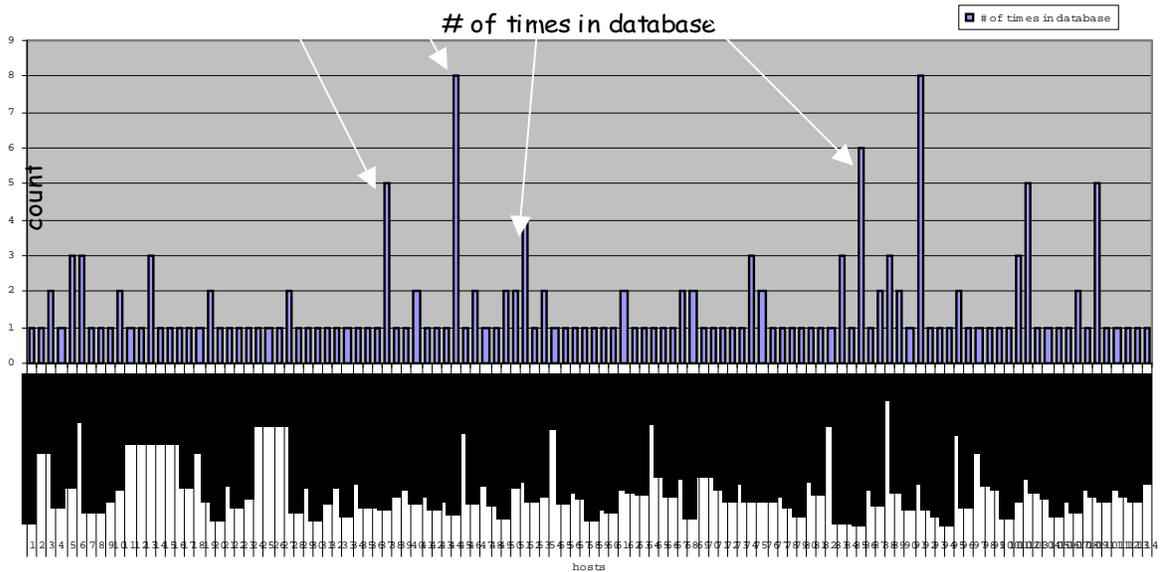

**Figure 16.** "The Usual Suspects" – Histograms of Frequently Compromised Computers

**Weakness Identification and Targeting:** Identifying the most frequent types of incidents in the organization's computer network can be critical when establishing future network security plans because it can help better target the security measures and identify weaknesses, leading to a reduced number of incidents. The SQL query is shown in Table 9.

```
SELECT types.name, count(incidentsid) AS cnt
FROM types, incidents
WHERE typeid=type
GROUP BY types.name ORDER BY cnt;
```
**Table 9.** Query to Identify Frequent Attacks

## 6. Summary

UCLog+ is a data management system for different forms of security data. UCLog+ pulls alerts and incidents into a centralized database that can be accessed via a web interface. The novel contribution of UCLog+ is the ability to also transparently incorporate raw data logs located across networks into the same system with the centralized database so pertinent information that otherwise would not be used can now be used in security analysis. The opposite approach -- to load raw data logs into a centralized database -- is not scalable, our UCLog+ approach is not only more scalable but also more reliable and more secure.

In this paper we used security data in the form of alerts, incidents, and raw data logs but other sources may be incorporated due to modular design. The preliminary results using UCLog+ we report here provide insight into attack trends on our network. Future work continues on developing UCLog+ so it can be shared via Internet download. When the code becomes stable we intend to transition to open source.